\documentclass[a4,12pt]{article}

\usepackage{amssymb}
\usepackage{latexsym}

\newcommand{\be}{\begin{equation}}
\newcommand{\ee}{\end{equation}}

\newcommand{\ie}{{\it i.e.}}
\newcommand{\eg}{{\it e.g.}}
\newcommand{\cf}{{\it c.f.}}

\newcommand{\sig}{\Sigma}
\newcommand{\csig}{\tilde{\Sigma}}

\newcommand{\Atot}{\mathcal{A}}
\newcommand{\A}{\mathcal{A_\infty}}
\newcommand{\Ag}{\A/\Gfo}

\newcommand{\Aia}{A^i_a}
\newcommand{\cAia}{\tilde{A}^i_a}
\newcommand{\cA}{\tilde{A}}

\newcommand{\comp}{\circ}

\newcommand{\g}[1][]{\gamma_{#1}}
\newcommand{\x}[1][]{x_{#1}}
\newcommand{\clp}{\ell}
\newcommand{\al}{\alpha}
\newcommand{\bet}{\beta}
\newcommand{\all}{\bar{\al}}
\newcommand{\bel}{\bar{\bet}}

\newcommand{\ra}{\rightarrow}

\newcommand{\snorm}[1]{\| #1 \|_{\infty}}
\newcommand{\lnorm}[1]{\| #1 \|_{2}}

\newcommand{\F}{\mathop{\mathrm{Fun}}}
\newcommand{\Fo}{\mathop{\mathrm{Fun_0}}}

\newcommand{\map}[3]{\mbox{$#1:#2 \rightarrow #3$}}

\newcommand{\D}{\mathcal{D}}
\newcommand{\G}{\mathcal{G}}

\newcommand{\Df}{\D_F}
\newcommand{\Gfo}{\mathcal{G}_{F}^0}
\newcommand{\Dfo}{\mathcal{D}_{F}^0}
\newcommand{\Di}{\D_{\infty}}

\newcommand{\Dio}{\D_{\infty}^0}
\newcommand{\Go}{\G_0}

\newcommand{\Hk}{H_{kin}}

\newcommand{\mcg}{G}
\newcommand{\sym}{S}

\newcommand{\rot}{[R]}

\newcommand{\drot}{R}

\newcommand{\R}{\Bbb{R}}
\newcommand{\Z}{\Bbb{Z}}
\newcommand{\C}{\Bbb{C}}

\newtheorem{theorem}{Theorem}
\newtheorem{lemma}{Lemma}
\newtheorem{conjecture}{Conjecture}

\newcommand{\Il}{\bar{I}}

\newcommand{\mapdown}[2]{\Big\downarrow
\rlap{$\vcenter{\hbox{$\scriptstyle#1$ \hspace{10mm} $#2$}}$}}

\newcommand{\bundle}[4]{\begin{array}[t]{c}
                #1\vspace{.5mm}\\ \mapdown{#2}{#4} \vspace{.75mm}\\ #3
		\end{array}} 

\newcommand{\fibrebundle}[6]{
		\mbox{$#1 \hspace{3mm}
		\stackrel{#4}{\longrightarrow} \bundle{#2}{#5}{#3}{#6}$}
}

\newcommand{\e}{\epsilon}

\newcommand{\calb}{\mathcal{B}}

\newcommand{\dec}{d}

\title{Existence of Spinorial States in Pure Loop Quantum Gravity}
\author{Matthias Arnsdorf\thanks{m.arnsdorf@ic.ac.uk} \\ \& \\
 Raquel S. Garcia\thanks{garciars@ic.ac.uk} \thanks{R.S.Garcia is
supported by a Beit Fellowship}
\\
\\
{\it
Blackett Laboratory,}\\
{\it Imperial College of Science Technology and Medicine,}\\
   {\it     London SW7 2BZ, United Kingdom.}  
}
\date{January 25, 1998}

\usepackage{graphics}

\begin{document}

\maketitle

\begin{abstract}
We demonstrate the existence of spinorial states in a kinematical theory of 
canonical quantum gravity without matter. This should be regarded as
evidence towards the conjecture that bound states with particle
properties appear in association with spatial regions of non-trivial
topology. In asymptotically trivial general relativity the momentum
constraint generates only a subgroup of the spatial
diffeomorphisms. The remaining diffeomorphisms give rise to the
mapping class group, which acts as a symmetry group on the phase
space. This action induces a unitary representation on the loop state
space of the Ashtekar formalism. Certain elements of
the diffeomorphism group can be regarded as asymptotic rotations of
space relative to its surroundings. We construct states that 
transform non-trivially under a $2\pi$-rotation: gravitational quantum
states with fractional spin.
\end{abstract}

\section{Introduction}

That states with fractional spin may exist in pure quantum gravity was
suggested as early as 1959 by Finkelstein and
Misner~\cite{finkelstein59}.  They believed that certain topological
properties of 3-manifolds may give rise to a notion of
spinoriality. Decades later, Friedman and Sorkin~\cite{friedman80}
showed that this possibility is implicit in a canonical formulation of
quantum gravity. The purpose of our paper is to provide evidence for
this conjecture by finding genuine spinorial states in the loop state
space of canonical quantum gravity.
 
Spinorial states in pure quantum gravity have been discussed in
connection with topological geons, bound states describing regions of
non-trivial spatial topology, whose origins lie in Wheeler's
work~\cite{wheeler55}.  It is therefore relevant to review the
mathematical framework which leads to the idea of a `particle of
topology'.  An underlying issue is the search for a quantum theory
given an appropriate classical limit, which is often a highly
non-trivial task. In particular, when the classical configuration
space of a system is not simply connected it is possible to construct
many quantum theories, each with its own physical properties.  This is
the case for Yang-Mills theories, where the different state spaces are
known as $\theta$-sectors. Our approach to quantisation sectors will
rest on the canonical view point, where the interplay between geometry
and group theory
\cite{balachandran91,isham83,giulini95} is best appreciated. 
Let us outline the main points.

A canonical formulation of a Yang-Mills theory is characterised by a
time-independent group of gauge transformations $\G$, which acts on
the phase space of the theory. This action can be viewed as an
internal symmetry transformation at each space point. The Gauss law
demands that we regard as equivalent all phase space points which are
related by the action of the connected component $\Go$ of $\G$,
while
invariance under all of $\G$ is
\emph{not} necessarily required. Hence the group $\G/\Go$ can act
non-trivially on the reduced phase space: the
space of orbits under the action of $\Go$.
When quantising we are lead
to consider a Hilbert space  carrying a unitary
representation $U$ of this quotient group.  Superselection sectors
arise when we demand that observables be invariant under the full
$\G$, which in the quantum theory  translates to the requirement that
operators should commute with all elements of $U$. In this way we
ensure that these operators do not mix states from different unitary
irreducible representations of $\G/\Go$, so that each of these
representations constitutes a separate quantum theory. If the internal
symmetry group of the theory is a non-Abelian simple Lie group and the
physical 3-space is simply connected, then it can
be shown that $\G/\Go = \Z$, where the parameter labelling the irreps
$e^{iz\theta}$ is the well-known $\theta$-angle from conventional
instanton theory.

The same applies to general relativity, where the analogues 
of the gauge transformations are the diffeomorphisms
and the canonical formalism demands only a \emph{subgroup} of 
diffeomorphisms to be redundancy transformations. We are left with the
non-trivial action of the mapping-class-group, which we will often call
$\mcg$. Interesting phenomena arise when one interprets this
group as the symmetry group of bound states. There has been extensive 
research on the structure of $\mcg$, here we just indicate the basic features.

We recall that in canonical approaches to gravity there is a preferred
3-manifold $\sig$ which represents physical space `at a given moment
in time'. It turns out that at a kinematical level one can
attribute particle properties to non-trivial pieces of topology, using
$\mcg$ as the asymptotic symmetry group. Let us make this statement a
little more concrete.  A region of space will give rise to a bound
state only when viewed in isolation by a distant observer.  We
capture this idea by demanding that $\sig$ be asymptotically
flat. There is a theorem according to which any such 3-manifold $\sig$
admits a unique decomposition --- up to diffeomorphism ---  into finitely
many basic constituents, that is:
\[
\sig = \R^3 \# P_1 \# \ldots \# P_n,
\]
The symbol $\#$ denotes the connected sum, which is the operation 
whereby two oriented 3-manifolds are glued
together, by removing an open ball from each and identifying the
resulting 2-sphere boundaries with an orientation-reversing
diffeomorphism. The basic elements $P_n$ in the decomposition are
\emph{prime manifolds}, which means that they cannot be further reduced
into the connected sum of other 3-manifolds apart from the 3-sphere
(note that $P \# S^3 = P$). For classifications and properties
of these manifolds we refer to~\cite{giulini93,hempel76}. We are
interested in their interpretation as elementary quantum geons, 
which finds support in another important theorem: the mapping class 
group of a manifold without handles,
\ie, without primes of the type $S^1 \times S^2$, is a
semi-direct product of three subgroups, each with its own
geometrical interpretation. Roughly we can describe them as: 
\setlength{\parskip}{0mm}
\begin{itemize}
\item \emph{permutations}, which act by exchanging identical primes;
\item \emph{internal diffeomorphisms}, which have support inside a given prime,
and hence can be viewed as its internal symmetries;
\item and \emph{slides}, which drag primes through other primes along a non-contractible
loop in $\sig$.
\end{itemize}
\setlength{\parskip}{5mm}
The interested reader will find details
in~\cite{sorkin95}. Among the internal diffeomorphisms, one of
particular interest is the $2\pi$ rotation, since it is associated with
the notion of spinoriality of a geon.  By composing $2\pi$ rotations
and exchanges one can investigate correlations between spin and
statistics of geons \cite{dowker96}. More generally, properties of topological 
geons arise via the representation theory of the mapping-class-group of the 
background 3-space $\sig$.  Although finite-dimensional representations have 
been studied and catalogued for a wealth of $3$-manifolds,  no attempts 
have been made so far to represent $G$ on a physical Hilbert space. 
One would nevertheless expect $\mcg (\sig)$ to induce an action on the state 
space in any canonical  formulation of quantum gravity based on $\sig$.

 What type of representation this will be, however, is not known. For
example it would be a priori conceivable that any proper gravitational
state space carries only a trivial representation of the mapping class
group. Luckily,
one needs not be so pessimistic, for we will show that the mapping
class group induces a non trivial action on the loop state space of
canonical quantum gravity, one of the recent candidates for a
gravitational Hilbert space. This is the second main theme of our
paper.

A reason why physical representations have not been considered so far is
that a geometrodynamical formulation of canonical gravity, where the
basic fields are Riemannian metrics on $\sig$, lacks a well defined
quantum counterpart at the kinematical level. In other words, 
even neglecting Hamiltonian evolution, it has not been possible to
construct a quantum state space rigorously. Over the years this
situation has greatly improved. With the introduction of the Ashtekar
variables \cite{ashtekar95}, based on connections and triads rather
than metrics, it was possible to overcome many of the technical
problems that had been hampering the definition of a kinematical state
space for quantum general relativity. A genuine Hilbert space was
finally introduced, the loop state space.  This is the space we will
use in our search for physical representations of the mapping class
group. In this paper we concentrate on just one aspect of the particle
nature of geons: spinoriality. We investigate whether a single prime
manifold, when used as the constant slice of canonical quantum
gravity, can lead to states with half-integer spin.

We have organised the material as follows. In section~\ref{Hk.section}
we define the quantum Hilbert
space where we will construct spinorial states. In section
\ref{spinorialman.section} we discuss in detail the definition of 
spinoriality in terms of diffeomorphisms of 3-manifolds. The results
from the first two sections are combined in section
\ref{spinorialstate.section}
 to obtain a condition for the existence of spinorial states. Then we
demonstrate that these condition are fulfilled for a particular prime,
the 3-torus. In this way we provide evidence for the original
conjecture. To conclude we indicate how our results can be generalised
and we suggest some avenues of future research.
Basic knowledge of fibre bundles and also of the Ashtekar formalism is
assumed. The references cover general background as well as more
specific topics.

\section{The kinematical Hilbert space}\label{Hk.section}

In this section we describe how the construction of the kinematical
Hilbert space used in loop quantum gravity can be adapted to the case
where space is no longer compact but asymptotically flat.
 The
restriction to a \emph{kinematical space} is due to the fact that a
satisfactory definition of a quantum Hamiltonian is still lacking and
presents a major obstacle in the construction of a complete theory of
quantum gravity. Nevertheless it is hoped that essential `kinematical'
features will carry over to a full theory as is often done in
discussions of `quantum geometry'.  Following Baez's
exposition~\cite{baez95}, we first introduce an auxiliary Hilbert
space, made of `cylindrical functions' of connections.  An action of
constraint operators can be defined rather naturally on this space,
which can be used to reduce it to obtain the final kinematical Hilbert
space.
  
\subsection{The auxiliary Hilbert space}

First we review the structure of the classical configuration space of
canonical gravity within the real Ashtekar formalism and the boundary
conditions imposed on the fields by the requirement of spatial
asymptotic flatness. In this formalism the space-time manifold
$\mathcal{M}$ is assumed to have the form $\mathcal{M} = \sig \times
\Bbb{R}$.  The condition of asymptotic flatness marks the topology of
$\sig$: it is required to have a regular end, \ie, a subset
$N$ (the neighbourhood of infinity) homeomorphic to $\R \times S^2$,
whose complement (the inner part of $\sig$) contains all the
distinctive topological features. In our laboratory interpretation the
cylinder $N$ represents the transition region between the isolated system
under study and the ambient universe.

Consider now the principal bundle of frames $P$ over $\sig$, which is
a trivial $SO(3)$ bundle\footnote
{
Note that we are using the \emph{real} Ashtekar variables~\cite{barbero94}. 
%In the literature it is often stated
%that the gauge group, even without inclusion of matter, is $SU(2)$,
%but this is misleading since in such an approach the group is
%represented unfaithfully (on $SU(2)$ spinors as opposed to triads),
%leaving us with a proper $SO(3)$ action
%{\it c.f.}~\cite{ashtekar91}.
}
(if, as we assume, $\sig$ is orientable).  Denote the space of smooth
connections on $P$ by $\Atot$.  Since our bundle is trivial, it has a
global cross section, which can be used to pull back the connections
to $so(3)$-valued one forms $\Aia$ on $\sig$, where $i$ denotes an
internal $so(3)$ index and $a$ a tensor index. If $\sig$ was compact
these connections would form the usual configuration space in the
Ashtekar formalism; for our non-compact $\sig$, the boundary condition
of asymptotic flatness implies that the connections `go to 0 at
infinity' at an appropriate rate. To make this precise we introduce a
radial coordinate system in the trivial region of $\sig$, defined with
respect to any metric on $\sig$ that is flat in this region. It
has been shown~\cite{ashtekar91,thiemann93} that in order to ensure
the equivalence of the geometrodynamic and connection formulation of
asymptotically trivial general relativity we need to impose the
following fall-off condition on our connections as $r
\rightarrow \infty$:
\be
\label{Asympt.eq}
\Aia = \frac{G^i_a}{r^2} + O(\frac{1}{r^3})
\ee
where $G^i_a$ is the leading order part of $\Aia$. Let us denote the
subspace of $\Atot$ whose elements satisfy this condition by $\A$.

Our discussion will be greatly simplified if instead of $\sig$ we use
its one-point compactification $\csig \equiv \sig \cup \infty$, which
is well-defined by the existence of a regular end in $\sig$, along
with the inclusion $\iota: \sig\hookrightarrow \csig$. 
In the
remainder of this section all quantities denoted by a tilde will
relate to $\csig$. 
  The fall off condition in
equation~(\ref{Asympt.eq}) allows us to extend all connections in $\A$
to infinity, \ie, for every $A \in \A$ there exists a
well-defined connection $\cA$ on $\csig$ with pull-back $\iota^*\cA =
A$. We can define these connections explicitly by doing a coordinate
transformation $r
\rightarrow v \equiv 1/r$ on the radial coordinates defined above. In these coordinates 
we have $\Aia = G^i_a v^2 + O(v^3)$, which tends smoothly to 0 as $v
\rightarrow 0$. Hence, taking $v = 0$ at infinity, 
we can set $\cAia = \Aia$ on $\iota(\sig)$ and $\cAia = 0$ at
$\infty$, which is a covariant, and hence well-defined, boundary
condition.

We are now ready to begin the construction of our Hilbert space. In a
finite-dimensional theory this would be a space of square-integrable
functions on the classical configuration space. But when considering
infinite-dimensional configuration spaces such as $\A$ this
construction is impeded by the lack of an analogue of the Lebesgue
measure.  This problem has been addressed in several papers
(see~\cite{ashtekar95} and references therein) and eventually solved
by defining a Hilbert space over an appropriate completion of $\A$,
where a measure \emph{can} be defined.  We now make use of the
compactification of $\sig$ to adapt the construction given by Baez
in~\cite{baez95} to the asymptotically flat case.  We will see later
that the use of $\csig$ is essential for the construction of
spinorial states.  The starting point towards a Hilbert space is to
select within the space of all functions on $\A$ a subset $\Fo(\A)$,
since this will make the definition of an inner-product possible.

So, let $\g:[0,1] \rightarrow \csig$ be a piecewise analytic
path\footnote{ The analyticity requirement is not strictly necessary
but is usually made for simplicity~\cite{baez95a}.  All our
topological considerations will be valid for the more general smooth
case.} in our compactified manifold.  Any connection $A \in \A$ gives
rise to a holonomy $H(\cA,\g)$, that is, to an $SO(3)$-equivariant map
between the fibres of $\tilde{P}$ over the endpoints of the path,
where $\tilde{P}$ is the principle bundle of frames over $\csig$. Since the
bundle $\tilde{P}$ is trivial we can identify its fibres through a global
cross-section and view the holonomy along a given path as an $SO(3)$
element. One then introduces, in analogy with the cylindrical
functions used in field theory, the algebra of functions of $\A$
generated by those of the form:
\[
\psi_{f,\Gamma}(A) = f(H(\cA,\g[1]),\ldots,H(\cA,\g[n]))
\]
where the set of paths $\Gamma \equiv \{\g[1],\ldots,\g[n]\}$ is an
embedded graph\footnote{
An \emph{embedded graph} is a finite set of paths
$\map{\g[i]}{[0,1]}{\csig}$ that are embeddings when restricted to
$(0,1)$ and intersect, if at all, only at their endpoints.
}
in $\csig$,
and
$f$ is a continuous function taking $SO(3)\times \cdots \times SO(3)$ to
$\C$. Notice that the same functional can be defined through infinitely many
pairs $f, \Gamma$. For example, the Wilson loop on some closed path
$\gamma$ is the same as the product of holonomy traces along the two 
subpaths between any two points in $\gamma$. Now let $\F(\A)$ be the 
completion of $\Fo(\A)$
in the sup norm (which is well-defined since $SO(3)^n$ is compact):
\[
\snorm{\psi} = \sup_{A \in \A}|\psi(A)|,
\]
where we have dropped the subscripts for ease of notation. 
We continue with the
construction of the kinematical Hilbert space by making
$\F(\A)$ into an $L^2$ space; this amounts to specifying how to
integrate functions. Here one faces the problem of constructing
appropriate measures, which as mentioned before is extremely difficult.
 A way round this problem is to introduce a so-called
`generalised measure', which is a bounded linear functional on $\F(\A)$.
 One such functional, the uniform generalised measure, can be obtained by
using the Haar measure on $SO(3)$ and defining:
\be\label{measure.eq}
\int_{\A} d\mu \ \psi_{f,\Gamma}
  \equiv \int_{SO(3)^n}f(g_1,\ldots,g_n)dg_1 \cdots dg_n,
\ee
where  the symbol $\int d\mu$  signifies a map
$\Fo(\A)\rightarrow
\C$. Note that the right hand side does not depend  on
$\Gamma$. More importantly it can be shown
(\cf~\cite{baez95,baez94a,baez94b,baez94c}) that if $\psi_{f,\Gamma} =
\psi'_{f',\Gamma'}$ then \mbox{$\int d\mu\, \psi = \int d\mu\,\psi'$}.
 Since $\Fo(\A)$ is dense in
$\F(\A)$, the functional in equation~(\ref{measure.eq}) extends
uniquely to a continuous linear functional on $\F(\A)$, enabling us to
define the norm:
\be
\lnorm{\psi} \equiv \left(\int_{\A}  d\mu\  |\psi|^2\right)^{1/2},
\ee
and the inner product
\be
\langle \psi_1 | \psi_2 \rangle = 
\int_{SO(3)^n} f^*_1(g_1,\ldots,g_n) f_2(g_1,\ldots,g_n)dg_1 \cdots dg_n,
\ee

\noindent where, if the original functions $f_1$ and $f_2$ have a 
different number of arguments, say $f_1 : SO(3)^m \rightarrow \C$ with
$m<n$, we trivially extend $f_1$ to a function on $SO(3)^n$ which
does not depend on the last $n\! -\! m$ arguments. The auxiliary
Hilbert space $L^2(\A)$ is the completion of $\F(\A)$ with respect to
this norm. If desired this space can also be regarded as an $L^2$
space defined with respect to a genuine measure on some completion
$\overline{\A}$ of $\A$. Since this will not be needed in the rest of
our discussion we leave the details to the references.

\subsection{Constraints and the kinematical Hilbert space}

It is well known that canonical general relativity is a theory of
constraints.  Classically these play the dual role of restricting the
phase space and generating physically important flows. In the quantum
theory however, following the Dirac procedure, this translates to the
sole condition that physical states have to be annihilated by operator
versions of the constraints.

A consequence of the extra $SO(3)$ redundancy introduced in the
passage from metrics to triads is that there are three types of
constraints in the Ashtekar formalism, as opposed to two in
geometrodynamics. As mentioned in the introduction we will ignore the
Hamiltonian constraint, which is in some sense responsible for
dynamics and time evolution.  This leaves us with the momentum
constraint $\mathcal{H}_i$, which classically generates diffeomorphisms 
of the 3-manifold $\sig$, and the Gauss constraint, 
which generates gauge transformations of the principal bundle where 
the connection fields are defined.  This geometrical interpretation of the 
constraints greatly
simplifies the passage to the quantum theory. Instead of looking
for operator equivalents of the classical constraints, with all the associated 
operator ordering and regularisation problems, we demand that physical states 
be invariant under the diffeomorphisms and gauge transformations generated by
the constraints.

To recognise the diffeomorphisms generated by the momentum constraint, 
we need to remember the asymptotic structure of the theory. Let us start by
defining two subgroups of the diffeomorphism group of $\sig$:
\begin{eqnarray}
\Di(\sig) &\equiv&  \{\mbox{diffeos in $\D(\sig)$ that preserve the
asymptotic structure of $\A$}\}
\nonumber\\
\Df(\sig) &\equiv& \{\mbox{diffeos in $\D(\Sigma)$ that are asymptotically
		trivial}\}
\nonumber
\end{eqnarray}
We denote their respective identity components by $\Dio$ and $\Dfo$.
For example, the diffeomorphism $\phi$ belongs to $\Dfo$ iff there
is a continuous curve of diffeos $t
\mapsto \Phi(t)$ with $\Phi(t)$ asymptotically trivial for all $t$,
such that $\Phi(0)=id_{\Sigma}$ and $\Phi(1)=\phi$.  We will also call
elements of $\Dfo$ small and the remaining elements of $\Df$ large
diffeomorphisms. Analogous definitions apply for the gauge transformations.

Since phase-space is made of asymptotically-flat connections, it would
be meaningless to consider diffeomorphisms that take us out of this
space. Therefore $\Di$ is the total group of diffeomorphisms acting on
our theory. But the momentum constraint generates only the smaller
group $\Dfo$.  To see this, recall that, in general, $\mathcal{H}_i$
determines flows in phase-space by smearing with vector fields on
$\sig$. Thus from a vector field $\xi$, one obtains the generating
functional $\mathcal{H}(\xi): [A,\pi]\mapsto \int
dx^3\xi^i\mathcal{H}_i(A,\pi)$, where $A$ and $\pi$ are the canonical
coordinates on the phase space.  It is easy to see that the flow
generated by a given $\mathcal{H}(\xi)$ is precisely that induced by
the one parameter subgroup of diffeomorphisms of $\sig$ associated
with the vector field $\xi$. Without further restrictions, this would
lead to all diffeomorphisms in $\Dio$.

What is special about our non-compact theory is that only
asymptotically trivial\footnote{The precise fall-off
condition will not be relevant to our topological considerations.}
 vector fields can be used in the smearing,
which clearly reduces the flows generated to $\Dfo$.  This requirement
is made to satisfy the necessary condition that the generating
functionals have well-defined functional derivatives, in the sense
that no boundary terms appear in their variations with respect to the
canonical variables.  In a similar way one can derive that the Gauss
constraint only generates the gauge transformations in the group
$\Gfo$.

Let us now define unitary actions of the total transformation groups
on the auxiliary Hilbert space $L^2(\A)$, which are induced naturally
from the action of gauge transformations and diffeomorphisms on the
space of connections $\A$.  First we consider the gauge
transformations.  In general they are defined to be any principal
automorphism of the underlying principal bundle (the base space is
kept fixed).  Any such morphism $g$, pulls back each connection $A$ on
$P$ to some other connection, thus defining a group action of
$\G$ on $\mathcal{A}$, which we denote: $A
\ra gA$. Since the bundle of frames is trivial, gauge transformations
can be identified with the set of $C^{\infty}$ maps
$\map{g}{\sig}{SO(3)}$. So in this case the familiar form of the gauge
transformations:
\[
A_a(x) \rightarrow g(x)A_a(x)g^{-1}(x) + g(x)\partial_a g^{-1}(x),
\]
can be interpreted globally as an automorphism of the space of
Lie-algebra valued connection one-forms on $\sig$.  
The subgroup $\G_{\infty}$ is simply the 
subset of maps from $\sig$ to $SO(3)$ that respect the boundary conditions of 
the connection.  The action of $\G_{\infty}$ on $\A$
induces a representation of $\G_{\infty}$ on $\Fo(\A)$ and by
extension also on $\F(\A)$ and $L^2(\A)$ given by:
\[
(U_g\varphi)(A) = \varphi(g^{-1}A),
\]
where $g \in \G_{\infty}$. The fact that $U_g$ is a unitary operator
follows directly from our definition of the inner product.

  A unitary action of $\Di$ is defined analogously. Diffeomorphisms of
$\sig$ lift to the frame bundle $P$, therefore inducing a
transformation of the connections denoted: $A \ra \phi A$ with $\phi
\in \Di$. In terms of the connection one-forms $\Aia$ on $\sig$ this
action is given by the ordinary pull-back by $\phi$.  We define the
 representation of $\Di$ on $L^2(\A)$ by:
\[
(U_{\phi}\varphi)(A) = \varphi(\phi^{-1}A).
\]
Unitarity also follows from the diffeomorphism invariance of the
generalised measure.
%shown in~\cite{baez94b}.
 
To implement the constraints we now have to demand that physical
states are left invariant by the actions of $\Gfo$ and $\Dfo$. Gauge
invariance is simply achieved by restricting to the subspace $L^2(\A/\Gfo)
\subset L^2(\A)$ of gauge invariant functions. A particular example of
such a state, which will be of importance later on is given by the
Wilson loop:
\be \label{state-inf}
\varphi(A) = tr[\rho(H(\cA,\clp))],
\ee
where $\clp$ is a loop in $\csig$ based at infinity and $\rho$ is any
representation of $SO(3)$. 
 A natural treatment of gauge-invariant
quantities was in fact one of the main motivations for using loop
variables.

In the case of the momentum constraint things get more complicated,
for it can be shown that there are no non-trivial diffeomorphism
invariant states in $L^2(\A/\Gfo)$. A similar situation arises even in
elementary quantum mechanics when considering constraint operators with a
continuous spectrum.  
These have non-normalisable eigenvectors, which means in particular
that the space of states satisfying the constraint --- eigenvectors with
eigenvalue 0 ---  is empty. 
This problem is overcome in two steps\footnote{
In the
series of steps  given towards the definition of $\Hk$, we have
omitted motivations and details. These come mainly from functional analysis
and the interested reader should consult the literature.
}~\cite{ashtekar95, arnsdorf96}.
First one considers distributional solutions to the momentum constraint.
 So instead of imposing the constraint
directly on $L^2(\A/\Gfo)$, one imposes it on the topological dual of
a dense subset thereof, which  
 can be taken to be $\F(\Ag)$. Secondly one selects a subspace $\mathcal{F}$ of
$\F(\Ag)'$ on which a natural definition of an inner product is possible.
The essential
point for us is that there is an induced action of $\Di$ on
$\F(\Ag)'$ and hence $\mathcal{F}$,
defined by:
\be \label{rep-diffeo}
(U'_{\phi}\psi')(\varphi) = \psi'(U^{\dag}_{\phi}\varphi),
\ee
where $U_\phi$ represents the action of the diffeomorphism $\phi$ on
$\F(\Ag)$, \mbox{$\varphi \in \F(\Ag)$} and $\psi' \in \F(\Ag)'$. Our final
reduced Hilbert space $\Hk$ is then given by the subspace of functionals
in $\mathcal{F}$
which satisfy:
\be
\psi'(U_{\phi}\varphi)= \psi'(\varphi) 
\ee
for all $\varphi$ and $\phi\in\Dfo$. Equivalently we can view
$\Hk$ as containing functionals on the quotient $\F(\Ag)/\Dfo$. A way to
construct such elements is via diffeomorphism group averages on
vectors in $\F(\Ag)$, see
\cite{thiemann97}. An alternative procedure through generalised measures 
can be found in \cite{baez94b}. It is shown that many non-trivial
elements actually exist.

\section{Spinorial manifolds} \label{spinorialman.section}

As we have discussed in the introduction there is a natural symmetry
group that appears in gauge theories, which is related to
configuration space topology and results in the emergence of the
well-known $\theta$-sectors. The analogous situation in general
relativity will be explored in more detail in this section. We will
introduce the mapping class group and show how it can be used to
define a notion of spin for the gravitational states.

We recall that the origin of $\theta$-sectors in gauge theories is a
discrepancy between all allowed gauge transformations (following Giulini we
denote these invariances) and those
generated by the constraints (redundancies). When considering general 
relativity the analogue of the gauge transformations are the
diffeomorphisms\footnote{In the Ashtekar variables we could have also
considered $\theta$-sectors arising as a result of $SO(3)$ gauge
invariance, but it can be shown that in our construction of the
Hilbert space there is no residual gauge action. This provides an
example that even if classically we have the necessary structure to
give rise to $\theta$-sectors (\cf~\cite{ashtekar91}), this does not
necessarily imply that these are realised in a quantum theory.}.  
Indeed we have seen that
only the group $\Dfo$ was required to be a redundancy group of our
physical system, while we were free to
define an action of the larger group $\Di$, which induces non-trivial
transformations of the rigid structure of $\sig$ at infinity.  After
construction of our final kinematical Hilbert space $\Hk$ we are thus
left with a residual group action of $\sym =
\Di(\sig)/\Dfo(\sig)$. Associated to $\sym$ is its discrete normal
subgroup $\mcg = \Df/\Dfo$: the mapping class group of
$\sig$.

In order to interpret the role of this group remember that usually
observables in general relativity are assumed to be invariant under
all diffeomorphisms. A way to proceed would be to regard $\sym$ as a
residual gauge group and use it to reduce the state space further,
either classically or quantum mechanically. We won't follow this
route, as it evidently rules out any interesting effects of the type
we are looking for. Another possibility is to consider $\sym$ as a
proper symmetry group rather than a gauge group.  This less
restrictive approach is the one to take when interested in detecting
superselection sectors. It is physically motivated by the observation
that our point at infinity does not represent an actual infinity,
instead it is a convenient model for the environment of our
system. This means that diffeomorphisms lying in $\sym$, which are
non-trivial at infinity correspond to genuine physical changes
relative to this environment.  Therefore we will let $\sym$ act
properly on our state space. The outcome of this proper action is what
we will be investigating. 
 For more discussion on
this point we refer to~\cite{giulini97, giulini94, landsman97}.
 
We will now take a closer look at these symmetries, especially their
relation to concepts of rotation and spinoriality.  It turns out that
in order to investigate the dependence of $\sym$ on the topology of
$\sig$ it is convenient to make use of the one-point compactification
$\csig$ of $\sig$ introduced earlier, since the natural subgroups of
diffeomorphisms in $\sig$ have analogues in the closed 3-manifold $\csig$,
which can be characterised in a very concise manner. So let us define
a few more subgroups of the respective diffeomorphism groups and
establish some homotopy equivalences. If $\phi$ is a diffeomorphism of
some manifold $M$ and $p$ a point in the manifold, then $\phi_*|{p}$
denotes the push-forward isomorphism from $T_p\, M$ onto
$T_{\phi(p)}\, M$. This general linear transformation will be said to
be in $SO(3)$ if its matrix representative with respect to some fixed
frame is in $SO(3)$.
\begin{eqnarray}
\Di(\csig) &\equiv&\{\mbox{$\phi$ in $\D(\csig)$ such
that $\phi(\infty) = \infty$ and $\phi_*|{\infty} \in
SO(3)$}\}
\nonumber\\
\Df(\csig) &\equiv&\{\mbox{$\phi$ in  $\D(\csig)$ such that
$\phi(\infty)=\infty$ and $\phi_*|{\infty}= 1$}\}
\end{eqnarray}
Each of these groups contains its identity component as a normal
subgroup. These definitions are well motivated since the extra motions
in $\Dio(\sig)$, over those in $\Dfo(\sig)$, are rigid 
 rotations at infinity, which are needed to define
spinoriality. Indeed, it can be shown that the spaces $\Di(\sig)$ and
$\Di(\csig)$ are homotopically equivalent, as are $\Df(\Sigma)$ and
$\Df(\csig)$~\cite{giulini94}.  In particular this implies that the
mapping class group $\mcg(\sig)$ is isomorphic to $\Df(\csig)/\Dfo(\csig) =
\pi_0(\Dfo) $, where the isomorphism is simply given by:
\be \label{isomorphism}
\map{\sigma}{\mcg(\csig)}{\mcg(\sig)}:[\phi] \mapsto [\phi|_{\sig}],
\ee
where $[\cdot]$ denotes an appropriate equivalence class of
diffeomorphisms. 
Our final
physical results will refer to the space $\sig$, but the above
isomorphism enable us to use $\csig$ as an equivalent ground where to
discuss the role of $\mcg$. From now on all diffeomorphism groups will
refer to $\csig$ unless stated otherwise.

To see that the mapping class group naturally leads to a concept of
spinoriality of gravitational states, we need a precise statement of
what we understand both by a rotation and by spinoriality. If we
adopted the general definition of a rotation as an \emph{element} of a
transformation group isomorphic to $SO(3)$, we would be precluding the
possibility of spinoriality.  Spinoriality means that a system
together with its environment is not returned to its initial state
after the system is rotated by $2\pi$. This can also be a macroscopic
property as 
is suggested by Dirac's string problem (\cf~\cite{penrose84}):
under certain boundary conditions, the $2\pi$-rotation of three
strings cannot be undone while their $4\pi$-rotation still can. The
key to allow for such effects is to look at rotations as continuous
sequences of transformations, \ie, as \emph{curves} in
$SO(3)$. Of course, what this implies is that we may be
\emph{actually} talking about $SU(2)$, the universal cover of $SO(3)$.

In our case we have to investigate how subgroups of the transformation groups 
$\sym$ and $\mcg$ are in correspondence to path classes in $SO(3)$. 
To this end we need to make use of the following principal fibre bundle:

\be \label{DF-Dinf-SO(3)}
\fibrebundle{\Df}{\Di}{SO(3)}{}{p}{p(\phi) \equiv \phi_*|_{\infty}}
\ee
and the associated fibration:

\be \label{G-S-SO(3)}
\fibrebundle{\Df/\Dfo = \mcg}{\sym}{SO(3)}{}{p}{}
\ee

The homotopy lifting theorem ensures that any loop in $SO(3)$ lifts to
a unique path through a given point in $\sym$. In particular, the end
point of a lift depends only on the homotopy class of the
original loop. Therefore the lift $\bar{\gamma}$ of any non-trivial
loop $\gamma$ in $SO(3)$ determines a homomorphism of $\pi_1(SO(3)) =
\Z_2$ into the fibre $\mcg$, given by \mbox{$\mu ([\gamma])=
[\bar{\gamma}(1)]_G$}. Spinoriality is decided according to the two
possible outcomes:
\setlength{\parskip}{0mm} 
\begin{enumerate}
\item[(i)] $\Z_2$ is
faithfully represented, that is $\bar{\gamma}(1)$ is not in the connected
component of $D_F$; 
\item[(ii)] $\Z_2$ is trivially represented, that is
$\bar{\gamma}(1)\in D_F^0$. 
\end{enumerate}
\setlength{\parskip}{5mm}
Which of the two occurs depends on the topology
of the manifold $\csig$, which will be respectively deemed as
spinorial and non-spinorial. In fact it can be shown (see
\eg~\cite{giulini93b,giulini94}) that for manifolds where (i)
holds\footnote{
The $\Z_2$ action is generated by $(-1,-1) \in G \times
SU(2)$, where the -1 on the right generates the centre of $SU(2)$, while
the -1 on the left belongs to some $\Z_2 \subset G$.
},
$\sym \cong\{\mcg \times SU(2)\}/\Z_2$; while for manifolds where (ii)
holds we have simply $\sym\cong \mcg \times SO(3)$. Hence a
\emph{spinorial manifold} is one whose associated symmetry group
contains $SU(2)$ but not $SO(3)$ as a subgroup.  The remaining
manifolds are non-spinorial. A complete classification according to
spinoriality has been established for all known 3-manifolds.

Now that we have a criterion for spinoriality, let us emphasise its 
geometrical content, so that we can visualise the elements of $\sym$ and $\mcg$
involved. We do so by explicitly lifting a non-trivial loop in $SO(3)$ to a
$2\pi$-rotation of a neighbourhood of infinity in $\csig$, which is a
rotation parallel to spheres~\cite{hendriks77} henceforth referred
to as a `prime
rotation'. Consider a
region $B$ around infinity small enough to be homeomorphic to $\R^3$
with associated radial coordinates. Consider also the 2-spheres $S_1$
and $S_2$ at values $r_1<r_2$ of the radial distance $r$ from infinity
in $B$. They bound two concentric balls $B_1\subset B_2$. Let $f:
B\rightarrow [0,1]$ be a continuous decreasing function of the radial
coordinate with
\be
f=\left\{\begin{array}{ll} 1  &\mbox{if $r<r_1$}\\
		       0   &\mbox{if $r>r_2$} \end{array}\right.
\ee
Then choose an axis in $B$ through the point at infinity and let 
$R(\theta)$ be the rotation of $B$ by $\theta$ around that axis.
The $2\pi$-rotation of $B_1$ gives a diffeomorphism $R$ of $\csig$
defined by:
\be
R \mathbf{x} =\left\{\begin{array}{ll}\mathbf{x} &\mbox{if 
$\mathbf{x}\in \csig - B$}\\
		R(2\pi f(r))\mathbf{x} &\mbox{if $\mathbf{x}\in B$}
                \end{array}\right.
\ee
In particular every point in $B_1$ is left invariant by $R$ so that
$R\in \D_F$.  The $2\pi$-rotation is moreover in $\Dio$, since
the curve of diffeos $R_{t}$ with
\be \label{hendriks.eq}
R_t \mathbf{x} = \left\{\begin{array}{ll} \mathbf{x} &\mbox{if 
		$\mathbf{x} \in \csig - B$}\\
	       R(t2\pi f(r))\mathbf{x}  &\mbox{if $\mathbf{x}\in B$}
	\end{array}\right.
\ee
has $R_t(\infty)=\infty$, $R_0=id_{\csig}$ and 
$R_1= R$.

The above defines a curve $R_t$ of diffeomorphisms from 
$id_{\csig}$ to the $2\pi$ rotation $R$, which projects\footnote{Note 
that any other choice of spheres $S_1$ and $S_2$, as well as any choice 
of rotation axis, leads upon projection to a homotopically equivalent loop and 
therefore defines the same element of $\mcg$.}  via 
$p$ to a non-trivial loop in $SO(3)$. All
the diffeomorphisms in the curve are representative elements of classes in 
$\sym$ and the endpoint $R$ is moreover an element of a class in $\mcg$. 
Thus $\Z_2$ is trivially represented if and only if $R \in \Dfo$. 
So in summary we can say that a 3-manifold is spinorial if a $2\pi$ 
rotation of a neighbourhood of infinity cannot be deformed to the identity 
along a curve of diffeomorphisms that leave the neighbourhood pointwise
invariant\footnote
{
Note that in topological considerations of
diffeomorphism groups we may replace frame-fixing diffeomorphisms with
those that leave an arbitrarily small neighbourhood of infinity
fixed~\cite{giulini93}.
}.

\section{Spinorial states}\label{spinorialstate.section}

We come now to the main part of the paper, where we put together
all the information gathered so far in order to derive sufficient
conditions for the existence of spinorial states.  In quantum theory
spinoriality is used to denote states with half integer angular
momentum, according to the representation theory of $SU(2)$.
Following this idea, one could try to define self-adjoint angular
momentum operators on the Hilbert space. This would involve finding
operator equivalents of classical ADM expressions and using these to
define spinoriality, as was done originally
in~\cite{friedman80}.  At the end of the day, one is really making use
of the rotations generated by the angular momentum operators. Indeed
we know that in quantum theory any weakly continuous unitary
representation of a one parameter group of symmetries will uniquely
define a self-adjoint operator. This motivates the definition of
spinoriality we will be using: spinorial states will be those that
change sign under the $2\pi$ rotation in some unitary representation
of the rotation group.

As we have argued in the previous section a $2\pi$ rotation is an
element of the symmetry group $\mcg$ that is obtained by lifting a
non-trivial loop in $SO(3)$. The endpoint $R$ of the lift is a
representative of the equivalence class $\rot$ of
diffeomorphisms. Whether these diffeomorphisms act on $\sig$ or
$\csig$ should be clear from the context. Because of the natural
isomorphism between $\mcg(\csig)$ and $\mcg(\sig)$ given by
equation~(\ref{isomorphism}), the same notion will be used in either case. 
We are interested
in the action of $\rot$ on $\Hk$. Since $\Hk$ is invariant under
$\Dfo$, equation~(\ref{rep-diffeo}) induces a well-defined
representation of $\sym$, and hence of its subgroup $\mcg$.  In
particular we can write:
\be \label{rep-rot}
\rot\, \psi'\,([\varphi]) = \psi'([U_R^{\dag}\varphi]),
\ee
where $[\varphi]$ is a class in $\F(\Ag)/\Dfo$ and $\psi'\in \Hk$. Then, a
sufficient condition for the existence of spinorial states is:
\[
\rot\,\psi'_1 =\psi' _2 \neq \psi'_1
\]
since then the state $\psi'_1- \psi'_2$ will be spinorial. Indeed, the
$4\pi$ rotation is always trivial, so that $\rot^2\,\psi'_1 =\psi'
_1$, and therefore $\rot\,(\psi'_1- \psi'_2) = - (\psi'_1- \psi'_2)$.
We see immediately that this is only possible if $\rot$ is not the
identity, \ie, if $R$ is not in $\Dfo$, which means that $\csig$ must
be spinorial. That spinoriality of $\csig$ is a necessary condition
for the existence of spinorial states was already established
in~\cite{friedman80}.  Whether it is also sufficient is an open
question, which can only be addressed in the presence of a concrete
physical Hilbert space. In the following we use the properties of
$\Hk$ to derive a topological criterion which ensures the existence of
spinorial states.

 Looking at equation~(\ref{rep-rot}), we see that if we find a vector
$[\varphi] \in \F(\Ag)/\Dfo$ such that $[U_{\drot}\varphi] \neq
[\varphi]$, then there will be a spinorial state in $\Hk$. In
particular, any functional in $\mathcal{F}$ that takes on different
values on these two equivalence classes will have the desired property.

In order to find a suitable vector in $\F(\Ag)/\Dfo$, we look for
loops in $\csig$ which are not $\Df$-ambient-isotopic\footnote{
 An
\emph{$(\Df-)$ ambient isotopy} of a manifold $M$ is a collection $\{I_t\}, t
\in [0,1]$ of homeomorphisms (diffeomorphisms in $\Df$) 
of $M$ onto itself such that the map $\map{I}{M \times
[0,1]}{M}:I(x,t) = I_t(x)$ is continuous. Two embeddings $e$, $e'$ in
$M$ are ambient isotopic if there exists an ambient isotopy $I$ of $M$
such that $I_0 = id_M$ and $I_1e = e'$.  } to their images under the
$2\pi$-rotation, since then cylindrical functions based on these loops
will transform non-trivially. We make this precise by looking at
states based on loops through infinity.  Consider the cylindrical
function $\varphi \in \F(\Ag)$ defined in
equation~(\ref{state-inf}). $R$ acts on $\varphi$ according to:
\[
 (U_{\drot}\varphi)(A)
= tr[\rho\, (H(R^{-1}\cA,\clp))]
\]
where we've used the symbol $R$ for the $2\pi$-rotation as a diffeomorphism in
$\sig$ and in $\csig$.
 From the definition of the holonomy it follows that:
\[
(U_{\drot}\varphi)(A) = 
 tr[\rho(H(\cA,R^{-1} \comp \clp))].
\]
 But we know that given two distinct loops $\clp_1$ and $\clp_2$,
 there is always a connection whose holonomy along $\clp_1$ is
 different from that along $\clp_2$. Now the elements in the
equivalence class
 $[\varphi]$ are related by the action of a small diffeomorphism.
Hence it follows that
if $\clp$
 and $R^{-1} \comp \clp$ are not related by an element of $\Dfo$, then
the states corresponding to these loops will lie in different classes and
 $[U_\drot \varphi] \neq [\varphi]$.  We conclude that the
 $2\pi$-rotation will be represented non-trivially on $\Hk$ provided
 there is a loop $\clp$ in $\csig$ such that no
 diffeomorphism in $\Dfo$ takes $\clp$ into its $2\pi$-rotation
 $R\comp\clp$; or in words, if two $R$-related loops $\clp$ and
 $R\comp\clp$ are not ambient isotopic relative to infinity.

Now it becomes clear why loops based at infinity are needed.  As explained
earlier we can choose $R$ to be any prime rotation around $\infty$:
they all define the same element of $\mcg$. Thus if a loop $\clp$
doesn't go through infinity we can define $R$ in a neighbourhood of
infinity disjoint from $\clp$. Therefore $R\comp\clp$ and $\clp$ 
are trivially ambient isotopic. This has an obvious physical
interpretation: in order to obtain spinorial effects the rotation of
our isolated system needs to be somehow communicated to the ambient
environment. This can only be done if the states themselves are highly
non local in the sense that they `extend to infinity'.

\subsection{Spinorial states and the fundamental group}

In this section we show that the prime manifold $\csig =T^3$ contains
loops not $\Df$-isotopic to their $2\pi$ rotation. As we have seen
this guarantees the existence of spinorial states in the Hilbert space
$\Hk$ associated with $T^3$. When discussing how these results can be
generalised we are lead to a condition that may be crucial for
spinoriality of states: the loops --- when considered as elements of
$\pi_1(\csig)$ ---  are not simply generated. As we will see this
condition ties in naturally with 
 the fact that all known spinorial primes have a
non-cyclic fundamental group.

\subsubsection{Existence of spinorial states in $T^3$}

The three-torus $T^3 = \R^3/\Z^3$ is known to be a spinorial
manifold.
 In the following we will consider a loop $\clp$ in $T^3$ and assume
 that it is $\Df$-isotopic to $R\comp\clp$, where $R$ is a
prime rotation in $T^3$. This will
lead to a contradiction, demonstrating that $\clp$ is spinorial. The
idea is to lift the loops and isotopies to $\R^3$, where with a little
work one can identify the structure with the Dirac string, which we
briefly describe.

Consider a solid object with three elastic strings attached, these
strings are tied to three separate posts. If we twist the object by
$2\pi$, around the axis of one of the strings say, then we obtain a
twisted set of strings referred to as the Dirac string. The twisting
cannot be undone by passing the strings over and around the object,
which provides an intuitive picture of spinoriality. We will
use this fact to
conclude that $R\comp\clp$ cannot be untwisted.

The Dirac string problem has a rigorous formulation in the language of
braid theory \cite{hansen89,birman74}, which we will be using below to
show how our problem reduces to it. First some definitions. A
\emph{braid} in $S^2$ is a system of $n$ embedded arcs \mbox{$\x
=\{\x[1],\ldots,\x[n]\}$} in $S^2 \times [0,1]$ such that: (i) each
arc $\x[i]$ intersects each intermediate sphere $S_t$ exactly once,
(ii) the arcs $\x[1],\ldots,\x[n]$ intersect each intermediate sphere
$S_t$ in exactly n different points. Two braids $\mathcal{B}_1$,
$\mathcal{B}_2$ in $S^2$ are said to be {\it s-isotopic} if there is
an ambient isotopy $I$ of \mbox{$S^2 \times [0,1]$} which takes
$\mathcal{B}_1$ to $\mathcal{B}_2$, such that the intermediate images
of the braid, $I_t\mathcal{B}_1$, during the isotopy define $S^2$-braids
and the endpoints of the braid are kept fixed.  The braid group
$B_k(S^2)$ is defined through the natural composition of s-isotopy
classes of braids with $k$ strings. A representative of the trivial
braid consists of $k$ radial strings from the inner to the outer
sphere. A simple presentation of the Dirac braid is obtained from the
trivial braid of three strings by a $2\pi$-rotation of the inner
sphere with respect to the outer sphere~\cite{hansen89}. The result is
a non-trivial braid, which is often denoted $\Delta$.

 $T^3$ can be presented as the unit cube centred around $0$ in
$\R^3$ with opposite sides identified. Let $0$ be the preferred point
called $\infty$ and consider two embedded loops $\alpha$, $\beta\,:
[0,1]\ra T^3$ at infinity, given by:
\[ 
\begin{array}{l} \al(s) = [h(s),0,0] \\
\\
\bet(s) = [0,0,h(s)] \end{array}
\]
where $h:[0,1]\rightarrow [-\frac{1}{2},\frac{1}{2}]$ is the function:
\[
h(s)=\left\{\begin{array}{lc} 
s & 0 \le  s \le \frac{1}{2}
\\
s -1 & \frac{1}{2} < s \le 1
\end{array}
\right.
\]

Our loop $\clp$ here is the composition $\alpha\beta$. We choose $R$
to be a prime rotation around the $z$-axis, centred at the origin and
leaving a ball of radius $\e$ invariant. $R$
maps $\bet$ into itself and maps $\al$ to a new loop which we denote
by $\al'$.
We have the following:
\begin{theorem}\label{t3.claim}
There is no diffeomorphism in $\Dfo(T^3)$ taking $\al\bet$ to $\al'\bet$.
\end{theorem}
{\bf Proof} We use the same notation for the elements involved in the
prime rotation as in section~\ref{spinorialman.section}, choosing
the radii of the balls $B_1$ and $B_2$ to be $\e$ and $\e'$ 
respectively. As anticipated, we proceed by contradiction. So suppose there is
a diffeomorphism in $\Dfo$ taking $\al\bet$ into $\al'\bet$. This implies the
existence of an isotopy $\map{I}{T^3\times [0,1]}{T^3}$ that leaves
the ball $B_1$ pointwise invariant and satisfies \mbox{$I(\al\bet,1) =
\al'\bet$} and \mbox{$I(x,0) = x$} for all $x \in T^3$.

\begin{figure}[!]
\centering\resizebox{12cm}{!}{\includegraphics{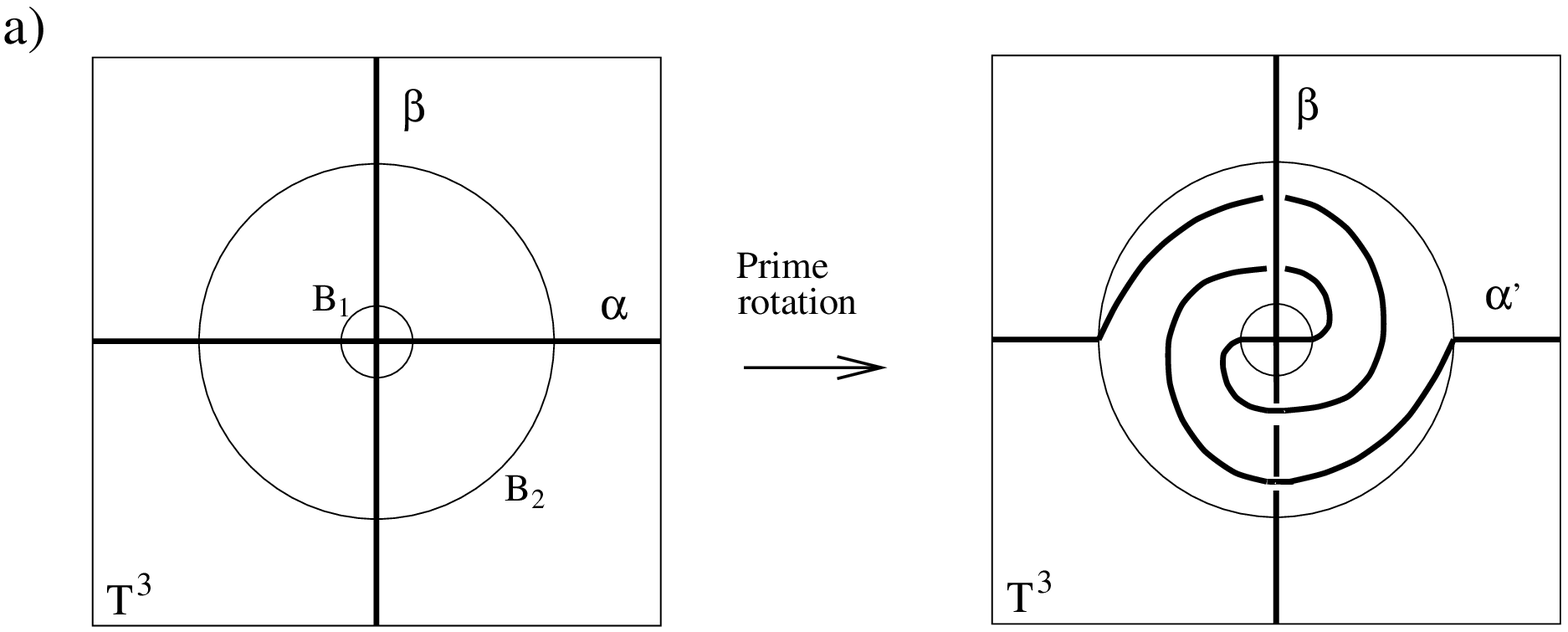}} \\
\vspace{0.55cm}
\centering\resizebox{12.2cm}{!}{\includegraphics{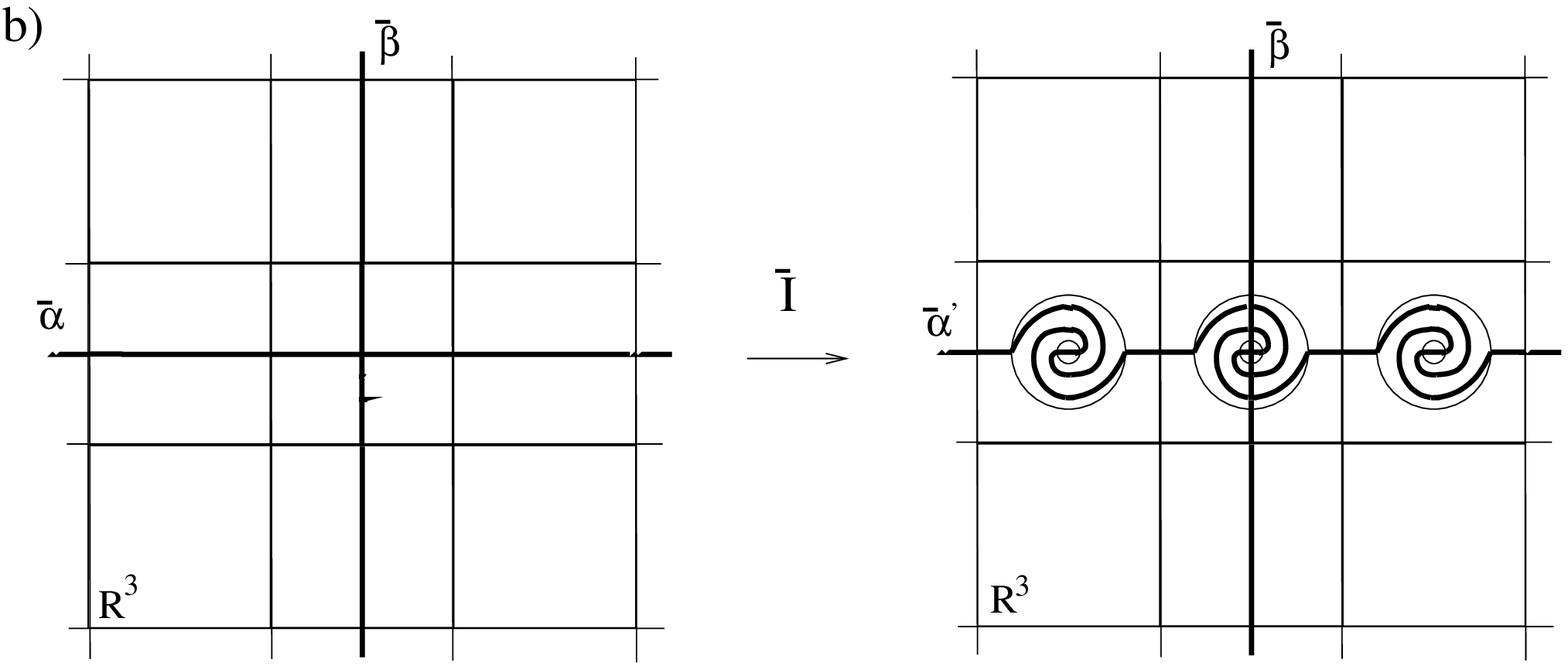}}\\
\vspace{0.5cm}
\centering\resizebox{12.2cm}{!}{\includegraphics{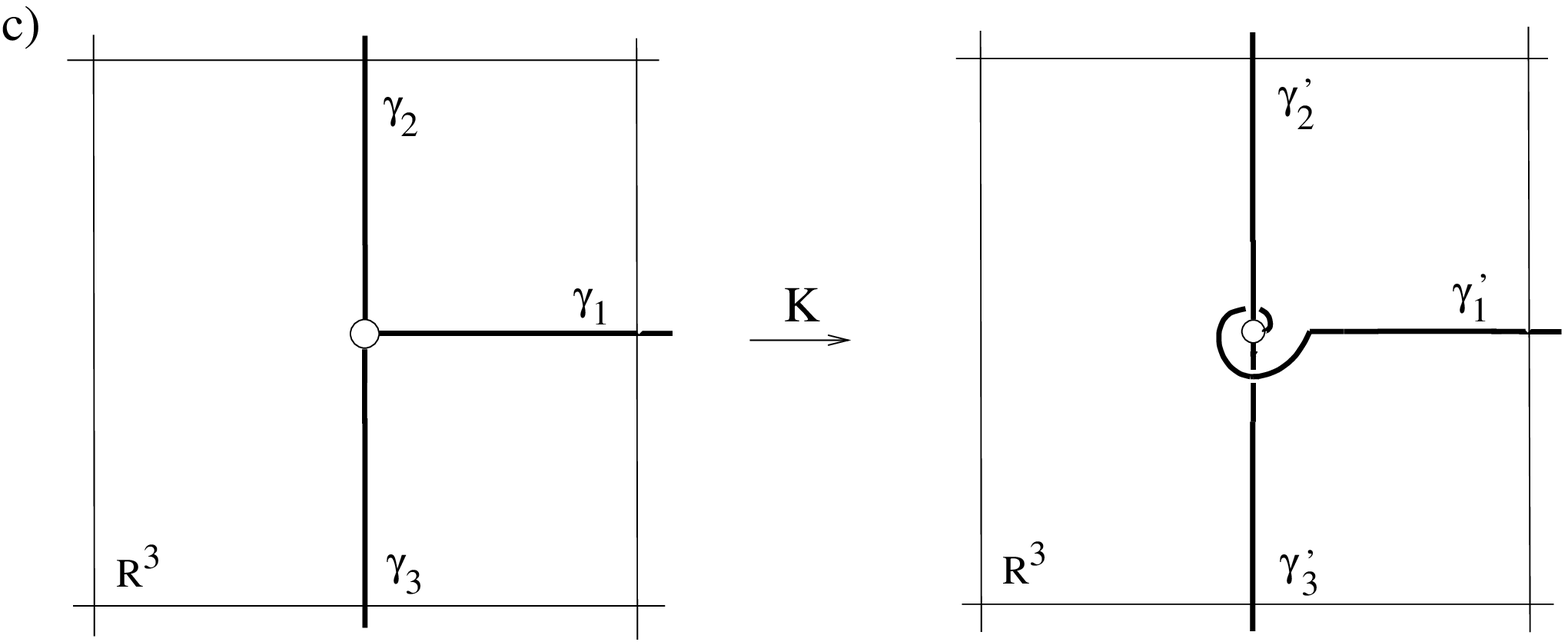}}

\caption{\small Figure a) depicts a schematic prime rotation about
the vertical  axis in 
$T^3$. Figure b) represents the lift $\Il$ to $R^3$ of the isotopy
$I$. Figure c) shows the action of the isotopy $K$ on the three
semi-infinite strings $\{\gamma_{i}\}$.  The restriction of the curves
to the ball $B_m$ of radius $m$ is identified with the ordinary Dirac
string $\Delta$.}\label{proof}

\end{figure}

We now lift the isotopy to $\R^3$, the universal covering of $T^3$, as
illustrated in figures~\ref{proof}a and~\ref{proof}b. Since $\Z^3$ acts
freely on $\R^3$ we have the following principal bundle:
\[\hspace{-4.2cm}
\fibrebundle{\Z^3}{\R^3}{\R^3/\Z^3}{}{p}{\quad p([x,y,z]) 
= [h(\dec(x)),h(\dec(y)),h(\dec(z))]}
\]
where $\dec(x)$ denotes the decimal part\footnote
{$d(x) \equiv x - \lfloor x \rfloor$, where $\lfloor x \rfloor$ is the
largest integer that is smaller than x.}
 of $x$. First, for each of
the loops $\al$, $\bet$ and $\al'$ there is a unique path: $\R \ra
\R^3$ through the origin, which we denote $\all$, $\bel$ or $\all'$,
defined by:
\[
p(\all(s+z)) = \al(s) \ \; \mathrm{and}\ \; \all(0) = 0
\]
with analogous expressions for $\bel$ and $\all'$. Here $s \in [0,1]$
and $z \in \Bbb{Z}$. These covering paths are just compositions of the
unique lifts of $\al$, $\bet$ and $\al'$ through different points of
the fibre $p^{-1}(\infty)$.
It is immediate that $\all$ is just the $x$-axis
and $\bel$ the $z$-axis. Secondly we lift the isotopy itself, using
the homotopy lifting property for fibre bundles (theorem 11.3 in
Steenrod~\cite{steenrod51}). It is easy to see that if the original
homotopy is an isotopy, then so is the lifted homotopy.  Thus the isotopy
$I$ lifts to an isotopy $\Il$ of $\R^3$ which covers $I$, that is:
\[
p(\Il(x,s)) = I(p(x),s)
\]
for $x \in \R^3$ and all $s \in [0,1]$. This isotopy has the property
that it leaves all points in $p^{-1}(\infty)$ fixed, as well
as a solid ball of radius $\e$ around each of these points.  Indeed, $I$
fixes all points in $B_1$ so that for each $x\in p^{-1}(B_1)$,
$\Il(x,s)$ lies in the fibre over $p(x)$ for all $s$. But since $\Il$
is continuous and the fibre is discrete $\Il$ must leave all points in
the fibre fixed. It is also easy to verify that:
\[
 \Il(\all,1) = \all',\hspace{5mm} \ \Il(\bel,1) = \bel
\]
and hence the lifted isotopy takes the cover of $\al$ into the cover
of $\al'$ and the cover of $\bet$ into itself. Since all points in
$\R^3$ with integer coordinates are fixed during the isotopy and since
$\Il$ is periodic, there is an integer radius $N$ such that for every
$n$ the images of the segments $\all_n= \all([n, n+1])$ and $\bel_n=
\bel([n, n+1])$ during the isotopy, stay within a ball of radius $N$
around $\all(n+1/2)$ and $\bel(n+1/2)$ respectively.

We may simplify $\all'$, by composing $\Il$ with another isotopy
$\Il'$ of $\R^3$ that reverses the prime rotation around every
point on $\all'$ with integer coordinates except $0$. This straightens
up all the segments $\all_n$ except for $\all_{-1}$ and $\all_1$.  We
conclude that there is an isotopy $K=\Il'\comp\Il$ of $\R^3$ that
takes $\all\bel$ onto a new path $\all''\bel$, where $\all''(t) =
\all'(t)$ for $|t| \le \frac{1}{2}$ and $\all''(t) = \all(t)$ for $|t|
> \frac{1}{2}$ (\cf\ figure~\ref{proof}c). In addition we are free to
impose: (i) $K$ leaves all points in $\R^3$ with integer coordinates
as well as the ball $B_1$ fixed; (ii) during the whole of $K$ the
curve $\all$ remains within $C_{\all}$ and $\bel$ remains within
$C_{\bel}$, where $C_{\all}$ and $C_{\bel}$ are solid cylinders of
radius $N$ around the $x$ and $z$-axis respectively.

The contradiction is now apparent since triviality of $\all''\bel$,
together with the properties of $K$, would imply triviality of the
Dirac string. To see this more explicitly, consider the following six
curves: $[\e,\infty)\ra \R^3$:
\[
\g[1](t) = \all(t), \quad \g[2](t) = \bel(t), \quad \g[3](t) = \bel(-t)
\]
\[
\g[1]'(t) = \all''(t), \quad \g[2]'(t) = \bel(t), \quad \g[3]'(t) = \bel(-t)
\]

Choose some $m \in \Z$, $m > N$. If we identify the shell in $\R^3$
between the two spheres of radius $\e$ and $m$ with the product $S^2
\times [0,1]$ then the restrictions of the above curves to the
interval $[\e,m]$ define two braids $\calb=\{x_i\}$ and
$\calb'=\{x'_i\}$ in $S^2$. In fact $\calb'$ is precisely the Dirac
braid. It follows that there is no s-isotopy taking the set of three
curves $\{\x[i]\}$ to the set $\{\x[i]'\}$. Notice however that the
intermediate strings in our isotopy $K$ need not define a genuine
$S^2$ braid, since they can leave the outer sphere $S_m$ of radius $m$. Thus
at a first glance, the action of the isotopy $K$ on the braid $\calb$
seems to be more general than s-isotopy. However, as in the case of
braids in $\R^2$ where s-isotopy and isotopy are
equivalent~\cite{artin47}, we can convince ourselves that the
properties of $K$ prevent it from using this extra freedom.

\begin{figure}[h]
\centering\resizebox{10cm}{!}{\includegraphics{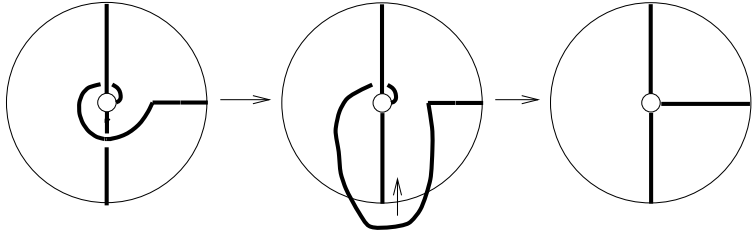}}
\caption{\small If considering curves of finite length, the Dirac twist can be
undone if no restrictions are imposed on the isotopy taking the
initial to the final configurations.} \label{undo}
\end{figure}

Indeed, the only way to relate two braids which, like $\calb$ and
$\calb'$, are not equivalent under s-isotopy is by pulling one of the
strings $x_i$ outside the sphere $S_m$ and over one of the other
two\footnote{ 
This move would undo the Dirac twist if one of the
curves $\g[i]$ had finite length and $K$ was not restricted by
condition (ii) above as shown in figure~\ref{undo}.
}.
But this is
not possible because of property (ii) above and because $m > N$
implies: \[ C_{\all} \cap C_{\bel} \cap
(\R^3- B_m) = \emptyset, 
\] 
where $B_m$ is the solid ball of radius $m$ centred at the origin.

We conclude that $K$ makes use of no additional generators to those in
$B(S^2)$.  It follows from non-triviality of the Dirac braid,
$\calb'$, that $K$, $\Il$ and $I$ do not exist $\Box$.

\subsubsection{Generalisations and conjectures}

\begin{figure}[h]
\centering\resizebox{62mm}{!}{\includegraphics{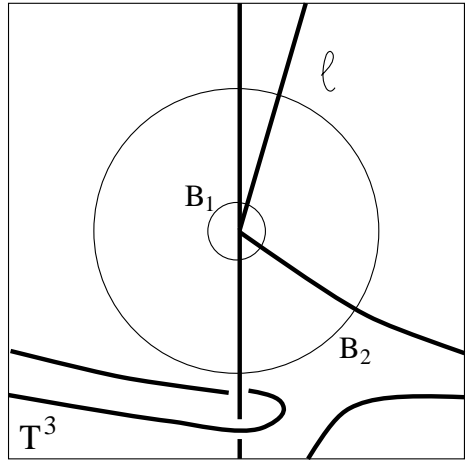}}
\caption{\small This is a picture of a singly-generated loop $\clp$ that is not 
ambient isotopic to $R\comp \clp$ through a curve of diffeos that leave $B_1$
invariant.}\label{thurston.fig}
\end{figure}

 In showing that the above isotopy $I$ does not exist an essential
point is that the loop classes $[\al]$ and $[\bet]$ are generated by
different elements of $\pi_1(T^3)$. We can also define spinorial
states using loops generated by the same element, provided their
isotopy class encodes the non-cyclicity of $\pi_1(T^3)$, as shown in
figure~\ref{thurston.fig}. On the other hand, given any generator
$[\g]$ of $\pi_1(\csig, \infty)$, if a loop $\clp$ is in the class
$[\g]^k$ for some integer $k$ and moreover $\clp$ is contained in a
single embedded solid 2-torus, then there is a diffeomorphism in
$\Dfo(\csig)$ taking $\clp$ to $\clp'= R\comp \clp$.

This suggests that a sufficient condition for the existence of
spinorial states may be established through the relationship between 
spinoriality of a prime and its fundamental group. We start by restating
the rather intuitive result that any prime with cyclic fundamental
group is non-spinorial. It depends on the Poincare conjecture \cite{hempel76}
according to which each closed, connected, simply connected $3$-manifold is 
homeomorphic to $S^3$.  As usual $\csig$ is any prime, $R$ any prime 
rotation around the preferred point $\infty$ of $\csig$ and $\Df$ the group of 
diffeomorphisms of $\csig$ that leave a ball around $\infty$ invariant.

\begin{lemma} \label{cyclic.lemma} Let $\clp$ be any loop through infinity in a prime $\csig$ with
$\pi_1(\csig)$ cyclic, then there is a diffeomorphism in $\Dfo(\csig)$ taking 
$\clp$ to $R\comp\clp$.
\end{lemma}
{\bf Proof}  It suffices to establish that any prime with cyclic fundamental group
is non-spinorial. The proof is divided in two cases:

a) If $\pi_1(\csig)= Z_p$, $\csig$ is non-spinorial since it has a finite fundamental
group with cyclic $2$-Sylow subgroup \cite{hendriks77}.

b) If $\pi_1(\csig) = Z$, then the Poincare conjecture implies that
$\csig$ is either $S^1\times S^2$ or the non-orientable handle,
usually denoted $S^1 \tilde{\times} S^2$, both of which are non-spinorial.

Note that in any case, we can always untwist the rotation of a
null-homotopic loop through infinity, by simply communicating the
$2\pi$-rotation outside a ball that contains the loop. This is as
expected, for such a loop can be regarded as lying in the regular end
of $\sig$ which means that it does not encode any of its non-trivial
topology. Hence there are many asymptotic non-spinorial states in any
given spinorial manifold.

A converse of the lemma above would be of great interest since
it would provide a sufficient condition for the existence of spinorial
states in $\Hk$. We know that all known orientable primes are spinorial if and only
if they have a non-cyclic fundamental group. For non-orientable
primes, such as $P^2\times S^1$, which is non-spinorial and has
non-cyclic $\pi_1$, this is not necessarily the case. This motivates:

\begin{conjecture}\label{noncyclic.conjecture} 
Let $\csig$ be an orientable prime.  Consider two embedded 
loops $\al$ and $\bet$ in $\csig$ that intersect only at $\infty$ and such 
that $[\al]$ and $[\bet]$ are distinct generators of $\pi_1(\csig)$. Then there is 
no diffeomorphism in $\Dfo(\csig)$ taking $\clp=\al\bet$ to $R\comp\clp =
\al'\bet'$.
\end{conjecture} 

The above result would clearly generalise to the case where $[\al]$ and
$[\bet]$ are not generators but are generated by distinct elements of
$\pi_1(\csig)$.
 A more important corollary would be the following:

\begin{conjecture}\label{spinorial.conjecture}
Every spinorial prime $\csig$ contains spinorial loops and
therefore $\Hk(\csig)$ contains spinorial states.  
\end{conjecture}

That this conjecture would follow from the previous one is clear, since by 
Lemma~\ref{cyclic.lemma} we know that any spinorial manifold has
non-cyclic fundamental group. Conjecture~\ref{noncyclic.conjecture} can be 
easily seen to hold when
$\csig$ is of the type $\R^3/H$, where $H$ is a discrete subgroup of
the affine group in 3 dimensions that acts freely and properly
discontinuously on $\R^3$. The proof follows from a slight
generalisation of the arguments leading to theorem~\ref{t3.claim}. More
general results, concerning all prime manifolds, will be given
elsewhere. They require further techniques of algebraic topology, in particular
of obstruction theory.
\begin{figure}
\centering\resizebox{14cm}{!}{\includegraphics{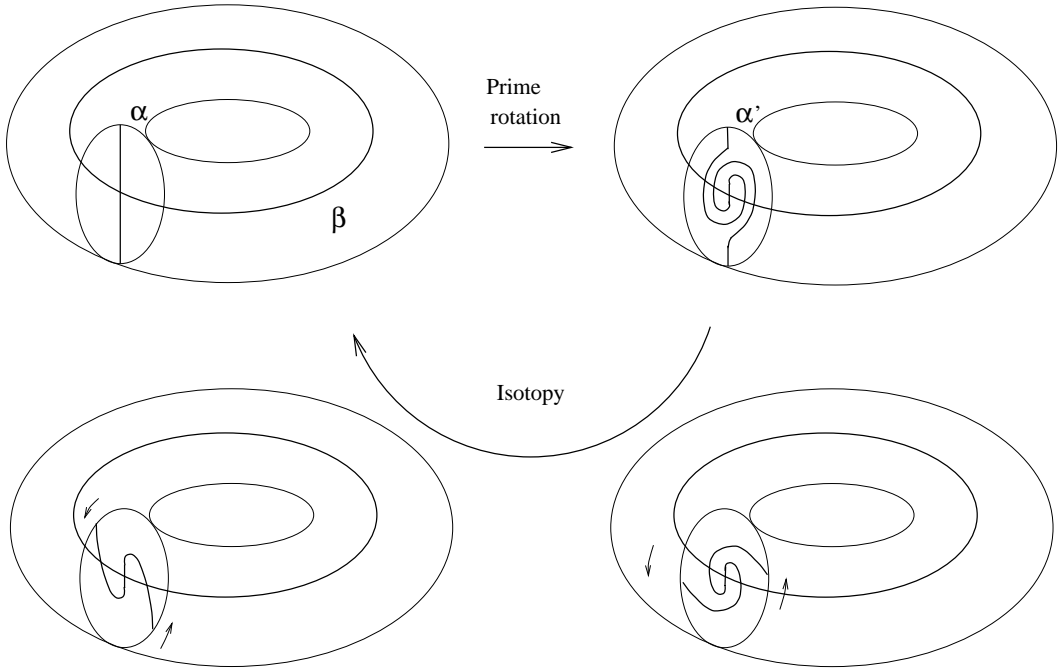}}
\caption{\small This is a picture of $P^2\times S^1$.
The disk represents $P^2$ so opposite points in its boundary are
identified.  Thus the loop $\al$ generates $\Z_2$ and the loop $\bet$
generates $Z$.  A $2\pi$-rotation turns $\al\bet$ into $\al'\bet$,
which can be isotoped back to the original loop by sliding the point
along the disk boundary.}\label{norient.fig}
\end{figure}

We conclude this section by illustrating explicitly how conjecture 
\ref{noncyclic.conjecture} can be refuted if orientability of 
$\csig$ is not required as a premise. The fundamental group
of the non-orientable prime $P^2\times S^1$ is $\Z_2\times \Z$. In
figure \ref{norient.fig} we start with the product $\clp$ of its two
distinct generators $\al$ and $\bet$, apply a prime rotation that
deforms $\clp$ into $\clp'$ and finally isotope back to $\clp$ without
moving the ball at infinity.

\section{Conclusion}

In the present report we have used the $2\pi$-rotation of an
$ADM$-slice relative to its surroundings to define genuine states with
half-integer spin in a kinematical Hilbert space for quantum
gravity. This reinforces the interpretation of the mapping class group
as an asymptotic symmetry group for bound states in a quantum 
theory of gravity.

Asymptotic symmetry transformations arise only in connection with
asymptotically flat spaces. However, the topological features of these
transformations can be analysed in the one-point compactifications of
the corresponding manifolds. This construction also allows a natural
definition of a Hilbert space for asymptotically trivial gravity,
which will be studied in more detail along with possible alternatives
in future.  A natural action of the mapping class group can be defined
on this Hilbert space and we have shown that it is precisely the
states through the added point at infinity which carry this
representation: localised states are left invariant.

Some elements of the diffeomorphism group can be regarded
as rotations of space with respect to its surroundings, leading to a notion of
spinoriality of a manifold and of the quantum states associated with it.
We have shown that spinorial states exist in a given manifold provided 
some loop based at infinity transforms non-trivially under the 
$2\pi$-rotation, in the sense that there is no ambient isotopy that leaves
invariant a neighbourhood of infinity and takes the original loop to the 
$2\pi$-rotated loop. Since in a non-spinorial manifold the $2\pi$-rotations
are trivial in the mapping-class-group, a necessary condition for the 
existence of spinorial loop states is spinoriality of the $ADM$-slice. 
We believe that this condition should also be sufficient.

In quantum field theory one attempts to classify particles according
to irreducible representations of appropriate symmetry groups.  In
analogy, when decomposing a given representation of the mapping class
group into irreducibles, one may hope that some of the quantum charges
labelling the resulting sectors admit an interpretation as particle
properties.  In the future we intend to address this question by
reducing the actions of the mapping-class-group on the loop state
space.  Although for our purposes in this paper it sufficed to look at
single primes, certain particle properties can only be described in
more general connected sums. For example, the distinction between
fermionic and bosonic representations of $\mcg$ depends on the
appearance of identical primes within a given connected sum. Hence we
expect manifolds with several primes to play an important role in our
search for multiplet structures.

Another interesting project would be to study the extent to which the 
asymptotic spatial 
symmetries that we have seen arise in a canonical formulation of quantum 
gravity, are present in the sum-over-histories formalism. As mentioned in
section \ref{Hk.section}, the Hamiltonian constraint remains one of the main unsolved 
problems in canonical quantum gravity. Perhaps by relaxing some of the
canonical structure, it would be possible to implement dynamics in a way 
that allows for topology change.  The canonical Hilbert space structure would 
break down at the critical time levels where topological transitions occur. These 
hypersurfaces would account for interactions and the particle properties 
would vary according to the topology of earlier and later slices.

\section*{Acknowledgements}

We are greatly indebted to Chris Isham for supplying the original
motivation to study topological geons and for countless discussions
thereafter. We are also very grateful to John Baez and Nico Giulini
for providing valuable comments on an earlier draft of our
report. Finally we would like to thank Fay Dowker for introducing us
to the topic of geons, Rafael Sorkin for his incisive remarks and Bill
Thurston and Ric Ancel for improving our intuition on three-manifold
topology.

\bibliography{references}
\end{document}